\documentclass[12pt]{article}
\usepackage[utf8]{inputenc}
\usepackage{graphicx}
\usepackage{caption}
\usepackage{subfigure}
\usepackage{amssymb}
\usepackage{amsmath}
\usepackage{wrapfig}
\usepackage{float}
\usepackage{geometry}
\usepackage{color}
\usepackage{cancel}
\usepackage{ulem}

\newcommand{\dd}{\text{d}} 
\newcommand{\artanh}{\mathop {\rm artanh}\nolimits }

\usepackage{amsthm}

\title{Description of electromagnetic fields in uniformly accelerated frame}
\author{Bart\l omiej B{\c a}k }
\begin{document}
\maketitle
\begin{abstract}
It is shown that Maxwell equations for electromagnetic fields generated by the uniformly accelerated charge could be reduced to the Laplace equation (in \L obaczewski geometry) for a single scalar potential. The full solution of this equation  is presented. 
\end{abstract}

\section{Introduction}
The  behaviour of electromagnetic field generated by the uniformly accelerated charge is still an interesting subject of studies for many researchers. The origins go back to 1909, when Max Born derived formulae for the monopole charge \cite{Born}. Then, the same result was independently rediscovered and studied by many other physicists (e.g.,  \cite{Bondi},  \cite{Drukey},  \cite{roh1960}, \cite{schott}, \cite{somm},  \cite{tur}). Nowadays, new approaches and comments also appear in literature (eg., \cite{kang2021},  \cite{Montesinos2023}, \cite{shiekh2021}), due to the fact that the discussion of radiating phenomena, related with this problem, is still a matter of intense debate. For this paper, the most important approach comes from  Z.Ya. Turakulov \cite{tur}, who proposed the geometrical way to rewrite the Maxwell equations in co-moving frame, where electromagnetic field stays static, and reduce to one Laplace equation (in \L obaczewski geometry) for a single scalar potential. Moreover, using a specific (bispherical)  coordinate system, this equation could be separated and, then, solved explicitly. More precisely, radial ODE is a special example of Gegenbauer equation \cite{derez}, whereas the second part is related to Laplace equation on the unit two-dimensional sphere, which leads to spherical harmonics.

To clarify the whole dissertation and make it self-consistent, in the second  section are defined two frames (laboratory and co-moving) as a special coordinate systems. In the  third section, there are presented elements of geometrical electrodynamics. The fourth section contains a procedure of solving it, and  in the fifth section are exposed expressions for electromagnetic potentials and fields in co-moving and laboratory frames.

\section{Laboratory and co-moving frames}

The starting point is a four dimensional manifold  with a flat Lorentzian metric tensor~$g$:
\begin{eqnarray}
    g = -\dd \tau^2 + \left(\dd \eta^1\right)^2+ \left(\dd \eta^2\right)^2 + \dd \zeta^2 \,, 
    \label{g1}
\end{eqnarray}
where $(\eta^{\mu})=(\tau,\eta^1,\eta^2,\zeta)$ are Cartesian coordinates, what is called as a \textit{laboratory frame}. 

The trajectory of the particle with constant acceleration in $\zeta$ direction is described by the hyperbola:
\begin{eqnarray}
    \zeta^2 - \tau^2 ={\rm const}>0 \, .
    \label{hyp}
\end{eqnarray}
Of course, this equation allows two arms of the hyperbola, therefore to disambiguate the trajectory, the range of the $\zeta$ coordinate has to be restricted to $\zeta>0$. It means, that in this analysis only half of the space is taken into account.\\

The \textit{co-moving (Fermi) frame}, associated with the accelerated particle, with  coordinate system $(x^{\mu})=(t,x^1,x^2,\xi)$ is defined via the following transformation~\cite{kijchrus1995},~ \cite{kij2008}:
\begin{eqnarray}
 \left\{
\begin{array}{cl}
     \tau\ =&\frac{1+a \xi}a\, \sinh \left(a t \right)    \, , \\
  \eta^1 =& x^1\, ,\\
  \eta^1 =& x^2\, ,\\
  \zeta =&  \frac{1+a\xi}a\,\cosh \left(a t \right)   \, , 
    
\end{array}
 \right. \label{trans}
\end{eqnarray}
where $a>0$ is an acceleration parameter, coordinate $\xi$ ``numerates'' hyperboloids and  $t$ is a proper time coordinate on each hyperboloid.  Due to the fact, that only half of the space is parametrized ($\zeta$ has to be positive), it implies that $\xi>-\frac 1a$. Then, the metric tensor \eqref{g1} takes form
\begin{eqnarray}
    g=-N^2\, \dd t^2 + \left(\dd x^1\right)^2+ \left(\dd x^2\right)^2 + \dd \xi^2  \, ,
    \label{g2}
\end{eqnarray} 
where 
\begin{eqnarray}
    N:= 1+a\xi 
    \label{lapse}
\end{eqnarray}
denotes the \textit{lapse function}. In the Fermi frame, the particle is located in the origin and its trajectory is given by $\xi(t)=0$. Importantly, when $a\rightarrow 0$ then $N\rightarrow 1$, and  coordinates $(x^{\mu})$ stay Cartesian. Indeed, the  metric \eqref{g2} in this limit obtains the standard form \eqref{g1}. 

It will be useful to introduce the following Euclidean metric $g_F$:
\begin{eqnarray}
    g_F:=g\big|_{t=const}= \left(\dd x^1\right)^2+ \left(\dd x^2\right)^2 + \dd \xi^2\, ,
    \label{gF}
\end{eqnarray}
which is the induced metric on hyperboloid $t={\rm const}$, and the associated Riemannian metric~$\gamma$:
\begin{eqnarray}
    \gamma := \frac{1}{N^2}\, g_{F}=  \frac{1}{(1+a\xi)^2}\,\left[ \left(\dd x^1\right)^2+ \left(\dd x^2\right)^2 + \dd \xi^2 \right]\, .
    \label{metgam}
\end{eqnarray}
Obviously, metric $\gamma$ is conformally flat (conformally equivalent to $g_F$). Following Turakulov \cite{tur}, the simplification of Maxwell equations is obtained via passing to  the bispherical coordinates $(\mu,\theta,\phi)$:
\begin{eqnarray}
\left\{
\begin{array}{cl}
    x^1  = &    \frac 1a \frac {\sin \theta \cos \phi}{\cosh \mu - \cos \theta}  \, , \\ 
  x^2  = &  \frac 1a \frac {\sin \theta \sin \phi}{\cosh \mu - \cos \theta} \, , \\ 
   \xi   = &   \frac 1a \frac {\sinh \mu}{\cosh \mu - \cos \theta} - \frac 1a  \,  .
   \end{array}
\right.
   \label{bisfer}
\end{eqnarray}
As before, the ranges of new coordinates have to be specified. Angle coordinates have standard domains: $\theta\in ]0,\pi[$, $\phi \in ]0,2\pi[$, whereas radial coordinate  $\mu>0$. The above transformation implies the following form of  metric $\gamma$ \eqref{metgam}:
\begin{eqnarray}
    \gamma&=&\frac{1}{a^2 \sinh^2\mu} \, \left( \dd \mu^2 + \dd\theta^2 +\sin^2\theta\,  \dd \phi^2 \right)\, ,
    \label{metgamL}
\end{eqnarray}
The Riemannian  manifold described by the metric $\gamma$  is isomorphic to the one-sheet hyperboloid in Minkowski space, what corresponds to \L obaczewski (hyperbolic) geometry,  and will naturally appear in the description of the electromagnetic field.

\section{Electrodynamics}

Electrodynamics on general metric background is a well-known topic in the literature, e.g., \cite{kij1994},  \cite{kijpod2003}, \cite{kijkosc2000},  \cite{koscdok},  \cite{landau} (see \textit{PROBLEM} p.256). Although, in this paper, the formulation of electrodynamics is restricted to metrics which have diagonal form and non-trivial lapse function (cf.~\eqref{g2}) and will be briefly presented below. 

The starting point is a 1-form of the electromagnetic potential $A$, which is used to define the Faraday 2-form $f$ as follows:
\begin{eqnarray}
  f=\dd A \Longrightarrow f_{\mu\nu} = \partial_{\mu}A_{\nu} - \partial_{\nu}A_{\mu}\, .
  \label{1form A}
\end{eqnarray}
The dual tensor density $\cal F$ is defined as a composition of a Hodge dual of $f$ and the so-called Weyl isomorphism ${\cal I}_W$ (the dual representation of a differential form):
\begin{equation*}
   {\cal F} =\left({\cal I}_W \circ  \star \right) f \ \Longrightarrow {\cal F}^{\mu\nu} = \frac 12\, \epsilon^{\mu\nu\alpha\beta}\, \left(\star f \right)_{\alpha\beta} =  \sqrt{|\det g|}\, f^{\mu\nu}\, ,
\end{equation*}
what corresponds to the simplest, vacuum, linear case.  $\epsilon^{\mu\nu\alpha\beta}$ denotes the Levi-Civita symbol, which is a tensor density (with a proper weight), and the convention is $\epsilon_{0123}=-\epsilon^{0123}= 1$. Hence, in geometrical unit system, where $G=c=\epsilon_0=\mu_0=1$ (see \cite{Gravitation}), the electric/magnetic induction is equal to the electric/magnetic field strength respectively. Here, the symbol $D$ describes the electric field, whereas $H$ denotes the magnetic field. Furthermore, all information about fields $D$ and $H$ is contained in components of Faraday 2-form $f$:
\begin{eqnarray} 
f_{kl} &=& \epsilon_{klm}  {\cal H}^m \ , \label{cfkl}\\
f_{0l}  &= & - N  D_l  \ , \label{cf0l} 
\end{eqnarray}
or, equivalently, in the dual $\cal F$:
\begin{eqnarray} 
{\cal F}^{kl} &=& - N \epsilon^{klm}\,  H_m \ , \label{cFkl}\\
{\cal F}^{0k} & = &  {\cal D}^k \ , \label{cF0l} 
\end{eqnarray}
where $N$ is a lapse function \eqref{lapse}, $\cal D$ and $\cal H$ are, respectively, electric and magnetic field densities related to the induced metric $g_F$ \eqref{gF}:
\begin{eqnarray}
     {\cal D} =\sqrt{\det g_F }\,  D \, ,  \qquad  {\cal H} = \sqrt{\det g_F }\,  H \, .
     \label{gestosci}
\end{eqnarray}
$\epsilon_{klm}$ denotes Levi-Civita symbol restricted to the spacial coordinates (which is also a tensor density with proper weight!), and convention is $\epsilon_{123}=\epsilon^{123}=1$. 

Maxwell equations in geometric manner are written as:
\begin{eqnarray}
    \dd f=0\, , \qquad \text{Div} {\cal F} = {\cal J}\ \Longrightarrow \ \partial_{[\alpha}f_{\mu\nu]}=0\, , \qquad \partial_{\nu}{\cal F}^{\mu\nu} = {\cal J}^{\mu}\, ,
\end{eqnarray}
where $\cal J$ is a vector density current. In this problem, the time component ${\cal J}^0$ is a  “suitable'' distribution  responsible for the point-like source (e.g., Dirac delta for the monopole charge), whereas other coefficients vanish. For electric and  magnetic fields, those equations take the following form:
\begin{eqnarray}
    \partial_k {\cal D}^k & = &{\cal J}^0\, , \label{cr1}\\
    \partial_k {\cal H}^k & = & 0 \, , \label{cr2} \\
    {\dot{\cal D}}^k & = &- \partial_l( N \epsilon^{lkm} H_m)\, , \\ 
    {\dot{\cal H}}^k & = &\partial_l ( N \epsilon^{lkm}     D_m) \, .
    \label{cr4}
\end{eqnarray}

The description  of a uniformly accelerated charged particle (with any multipole character) will be formulated as a stationary solution of Maxwell equations in the co-moved frame, generated by the point-like source. It is assumed, that the fields $D$, $H$ have to be static what implies, with equation \eqref{cr2}, that magnetic field $H$ is constant. Physically, the static charge has not produced any magnetic field, therefore, $H$ vanishes.  Moreover, equation \eqref{cr4} is needed for existence of the scalar function (potential) $U$\footnote{The first cohomology space $H^1$ in the corresponding three-dimensional subspace is trivial.}, defined as:
\begin{eqnarray}
    \partial_k U := N\, D_{k}\, .
    \label{def U}
\end{eqnarray}
According to \eqref{1form A} and \eqref{cf0l}, $U$ is the first  component of electromagnetic potential 1-form $A$ in Fermi frame:
\begin{eqnarray}
    A = U\, \dd t\, ,
    \label{udt}
\end{eqnarray}
Using \eqref{g2}, \eqref{metgam} and \eqref{gestosci} it is a matter of simple calculations to prove that the first Maxwell equation \eqref{cr1} is equivalent to the Laplace equation for potential $U$:
\begin{eqnarray}
     \partial_k {\cal D}^k = \sqrt{\det \gamma}\, \Delta_{\gamma} U = 0\, ,
     \label{laplaceeq}
\end{eqnarray}
where the symbol $\Delta_{\gamma}$ signs the Laplace-Beltrami operator for the metric $\gamma$. Of course, it holds everywhere except one point -- the origin of the co-moving frame which is also the support of the current's distribution, where the particle is located.  Therefore, the problem of uniformly accelerated  charge is, like in the standard static case, described by the Laplace equation (or, more precisely, by the Poisson equation with point-like sources), but for different Riemannian geometry. Precisely, the standard  static case corresponds to the Euclidean geometry, whereas uniformly accelerated frame to the \L obaczewski geometry.

\section{Solution of the Laplace equation}

The solution of the Laplace equation \eqref{laplaceeq} is made in few steps which, due to the complexity of this problem, are explicitly presented below.\\

Firstly, the equation \eqref{laplaceeq} has to be written in bispherical coordinates $(\mu,\theta,\phi)$, then:
\begin{eqnarray}
   \sinh\mu \, \partial_{\mu}\left( \frac{1}{\sinh \mu}\, \partial_{\mu} U\right) + \stackrel{\circ}{\Delta} U =0\, ,
   \label{eq0}
\end{eqnarray}
where $\stackrel{\circ}{\Delta}$ denotes the Laplace-Beltrami operator on the unit sphere $\mathbb{S}_2$ with coordinates $(\theta,\phi)$. The solution $U$ of the above equation is assumed to be a product of an eigenvector  $Y(\theta,\phi)$  of the operator $\stackrel{\circ}{\Delta}$:
\begin{eqnarray}
   \stackrel{\circ}{\Delta} Y_l(\theta,\phi) = -l(l+1)  Y_l(\theta,\phi) \, ,
   \label{eqsph}
\end{eqnarray}
which are called as \textit{spherical harmonics}, and the “radial part'' $R(\mu)$:
\begin{eqnarray}
    U(\mu,\theta,\phi) := R(\mu) Y(\theta,\phi)\, .
    \label{potU}
\end{eqnarray}
Hence, the equation \eqref{eq0} takes the following form:
\begin{eqnarray}
    \sinh\mu\,  \partial_{\mu}\left( \frac{1}{\sinh \mu}\, \partial_{\mu}R_l(\mu)  \right) -l(l+1) R_l(\mu) = 0\, ,
\end{eqnarray}
where $l\geq0$ numerates solutions with respect to the multipole (harmonic) expansion. Finally, taking 
\begin{eqnarray}
R_l(\mu)=\frac{1}{\sinh^l\mu}\, G_l(\coth \mu)
\label{podst}
\end{eqnarray}
and defining a new variable
\begin{eqnarray}
    z:=\coth \mu\, , 
    \label{defz}
\end{eqnarray}
where $z>1$ (because $\mu>0$), this equation corresponds to the special case of the Gegenbauer (hypergeometric) equation \cite{derez}:
 \begin{eqnarray}
       \left[(1-z^2)\, \partial_z^2-2(1+\alpha)z\,\partial_z +  \lambda^2- \left(\alpha+\frac12\right)^2\right] G_{\alpha,\lambda}(z) = 0\,,
       \label{gege}
    \end{eqnarray}
with parameters $\alpha=l+\frac 12$ and $\lambda=1$. Hence:
\begin{eqnarray}
 \left[ (1-z^2)\, \partial_z^2-(2l+ 3)z\,\partial_z +  1- \left(l+1\right)^2\right] G_l(z) = 0\, .
    \label{rowGeg}
\end{eqnarray}
Surprisingly, solutions $G_l(z)$ of the equation \eqref{rowGeg} are expressed by elementary functions, although  they split into two cases: $l=0$:
\begin{eqnarray}
    G_0(z)&=& \alpha_0 + \beta_0\left(1-\frac{z}{\sqrt{z^2-1}} \right)\, ,
    \label{l0}
    \end{eqnarray}
    and $l\geq1$:
    \begin{eqnarray} 
    G_l(z)&=&\alpha_l\, Q_l(z)  + \beta_l\, Z_l(z)\, , 
    \label{l1}
\end{eqnarray}
where $\alpha_0, \beta_0, \alpha_l, \beta_l$ are real constants, whereas $Q_l(z), Z_l(z)$ are functions  presented below:
\begin{eqnarray}
\begin{array}{c|c|c}
    l & Q_{l}(z) & Z_{l}(z) \\
    \hline
    1 & \frac{z \sqrt{z^2-1} - \log\left(z+ \sqrt{z^2-1} \right)}{\left(z^2-1\right)^{3/2}} & \frac{1}{\left(z^2-1\right)^{3/2}} \\
    2 &\frac{-(z^2+2)\,\sqrt{z^2-1}+3z\,  \log\left(z+ \sqrt{z^2-1} \right)}{\left(z^2-1\right)^{5/2}}  & -\frac{3z}{\left(z^2-1\right)^{5/2}}\\
    3 & \frac{-\left(13 +2z^2\right)z\, \sqrt{z^2-1} +3 \left(4 z^2+1\right) \,  \log\left(z+ \sqrt{z^2-1} \right) }{2 \left(z^2-1\right)^{7/2}} & \frac{3 \left(4 z^2+1\right)}{\left(z^2-1\right)^{7/2}} \\
    4&\frac{-\left(6 z^4+83 z^2+16\right) \sqrt{z^2-1}+15 z \left(4 z^2+3\right) \, \log\left(z+ \sqrt{z^2-1} \right)}{6 \left(z^2-1\right)^{9/2}} & -\frac{15 z \left(4 z^2+3\right)}{\left(z^2-1\right)^{9/2}}
\end{array}
\label{tab}
\end{eqnarray}
More details about Gegenbauer equation \eqref{rowGeg} are presented in Appendix~\ref{gegapp}.

\section{Potentials and fields}
The 1-form of electromagnetic potential $A$ is a very useful object in this description, due to the fact that in co-moving (Fermi) frame its time component $U$ \eqref{udt} is a solution of Laplace equation \eqref{laplaceeq} and it determines electric and magnetic fields via Faraday tensor $f$ \eqref{1form A}. Moreover, as a geometrical object, it could be easily transformed to any other coordinate system or reference frame. There were distinguished two special frames: laboratory and co-moving, therefore, it will be advantageous to present potentials and fields in these two cases what is done in the sequel.

\subsection{Co-moving (Fermi) frame}
\label{potfermiframe}

Using formula \eqref{podst} one could easily compute the explicit form of the radial function~$R$  and, multiplying by a proper spherical harmonic $Y$,  write the whole potential $U$~\eqref{potU} in bispherical coordinates $(\mu,\theta,\phi)$. Unfortunately, this coordinate system is rather a mathematical trick to simplify calculations than a physical object. To write potentials and fields in a useful way, it is necessary to develop appropriate coordinates corresponding to the symmetry of the problem. The natural choice is  the cylindrical coordinate system  $(r,\phi,\xi)$ (related with Cartesian coordinates $(x^{\mu})$ in co-moving frame!) due to the fact, that transformation between laboratory and co-moving frame change only time and axial coordinate -- see \eqref{trans} -- what is a consequence of a hyperbolic motion \eqref{hyp}. Therefore,  the bispherical coordinates $(\mu,\theta,\phi)$ have to be transformed to Cartesian coordinates $(x^1,x^2,\xi)$ (cf. \eqref{bisfer}) and then, implemented the standard polar coordinates:
\begin{eqnarray}
\left\{
\begin{array}{cl}
    x^1  = &    r \cos\phi  \, , \\ 
  x^2  = &  r \sin \phi \, .
\end{array}
\right.
   \label{cylin}
\end{eqnarray}
The potential in Fermi frame in cylindrical coordinates $U_F(r,\phi,\xi)$ \eqref{potU} is equal to the product of radial part represented by Gegenbauer function $G(\coth \mu)$ \eqref{podst} and spherical harmonics $Y(\theta,\phi)$, but written in new coordinate system. Hence:
\begin{eqnarray}
    U_{F}(r,\phi,\xi) := R(\mu(r,\xi))\,  Y(\theta(r,\xi),\phi)\, .
    \label{UF}
\end{eqnarray}
In radial function $R(\mu)$, the variable $\mu$ appears only by hyperbolic functions $\sinh \mu$ and $\cosh\mu$, therefore, it is sufficient to write formulae only for them:
\begin{eqnarray}
\sinh \mu &=&\frac{\Gamma^2-\Phi^2}{2\Phi \Gamma} \, , \\
\cosh \mu &=& \frac{\Gamma^2+\Phi^2}{2\Phi \Gamma}\, ,
\end{eqnarray}
where
\begin{eqnarray}
       \Phi &:=&\sqrt{a^2\left(r^2+\xi^2\right)}\,, 
       \label{Phi}\\
       \Gamma&:=&\sqrt{ a^2 (r^2 + \xi^2)+4(a \xi+1)}\, .\label{Gamma}
     \end{eqnarray}
Transformation between bispherical and cylindrical systems “preserves'' only the azimuthal angle $\phi$. Furthermore, in the subsequent sections, only harmonics for $l=0$ and $l=1$ will be used:
\begin{eqnarray}
   \left\{ \begin{array}{rl}
        {  Y}_0 (\theta,\phi)=&1\, , 
   \end{array} \right. \quad
   \left\{ \begin{array}{rl} 
   { Y}_x (\theta,\phi)=& \sin \theta \cos\phi  \, ,  \\
   {  Y}_y (\theta,\phi)=& \sin \theta \sin\phi  \, ,  \\
    {  Y}_z(\theta,\phi)=& \cos \theta \, .
   \end{array} \right.  
      \label{harmoniki}
\end{eqnarray} 
Accordingly, there are written, as above, transformation formulae only for trigonometry functions:
\begin{eqnarray}
    \cos \theta &=&\frac{a^2(r^2+\xi^2) + 2 a \xi }{\Phi \Gamma} =\frac{\Gamma^2+\Phi^2 -4 }{2\Phi \Gamma}  \, , 
    \label{costh}\\
    \sin \theta &=& \frac{2 a r}{\Phi \Gamma}\,.
    \label{sinth}
\end{eqnarray}
For simplicity, the “radial'' function $R_F$ in cylindrical coordinates is defined as:
\begin{eqnarray}
    R_{F}(r,\xi):=R(\mu(r,\xi))\, ,
    \label{RF}
\end{eqnarray}
where letter $F$ highlights the fact that those functions are written in Fermi frame. For $l=0,1$ the radial functions are written explicitly below:
\begin{eqnarray}
    R_{F,0} &=&\alpha_0\, \left[1- \frac{ a^2\left(r^2+  \xi^2\right) + 2( a \xi +1)}{\Phi\Gamma} \right]   + \beta_0 =  - \alpha_0\,  \frac{\left(\Gamma-\Phi\right)^2}{2\Phi\Gamma}   + \beta_0\, , \label{u10}\\
    R_{F,1}&=&  \frac{4(a\xi+1)^2\, \left[ \alpha_1 -2 \beta_1\,  \artanh\left(\frac{\Phi}{\Gamma} \right)  \right]}{ {\Phi^2 \Gamma^2}} +\frac{\beta_1\,\left[ a^2\left(r^2+  \xi^2\right) + 2( a \xi +1)\right]}{{\Phi\Gamma}} =\nonumber \\
    &=& \frac{(\Gamma^2-\Phi^2)^2}{4\Phi^2\Gamma^2}\, \left[ \alpha_1 - \beta_1\, \log\left(\frac{\Gamma+\Phi}{\Gamma-\phi} \right)  \right] + \beta_1\, \frac{\Gamma^2+\Phi^2}{2\Phi\Gamma} \, . \label{u11} 
\end{eqnarray}
The value of real constants $\alpha_0, \beta_0, \alpha_1, \beta_1$ depends on assumed boundary conditions.  As it was mentioned before, in co-moving frame appears only the electric field $D$ derived via \eqref{def U}. In Fermi frame with cylindrical coordinates, this field will be denoted as $D_F$. 

For $l=0$  is only one solution $D_{F,0}$ with two non-vanishing components:
\begin{eqnarray*}
    D_{F,0}^r &=& \frac{8\alpha_0\, a^2 r (a\xi+1) }{\Phi^3\Gamma^3} = \frac{2 \alpha_0\, a^2 r \left( \Gamma^2-\Phi^2 \right)  }{\Phi^3 \Gamma^3}\, , \\
    D_{F,0}^{\xi} &=&\frac{4\alpha_0 \, a   \left[a^2\left(\xi^2-r^2\right) + 2a\xi \right]}{\Phi^3\Gamma^3}= -\frac{\alpha_0\, a \left[4\left( \Gamma^2+ \Phi^2\right)-\left( \Gamma^2- \Phi^2\right)^2\right]}{2\Phi^3\Gamma^3} \, .
     \end{eqnarray*}  
For $l=1$ the situation is much more complicated due to the appearance of projections on three, non-trivial spherical harmonics. Cylindrical symmetry implies that, the full solution could be split into two cases: parallel to direction of movement (designed by $D_{\parallel}$) related with $Y_z$, and perpendicular  ($D_{\perp}$) related with $Y_x$ and $Y_y$~(cf. \eqref{harmoniki}). Hence: 
\begin{eqnarray}
    D^r_{F,\parallel}&=& \frac{a^2 r\, \left(\Gamma^2-\Phi^2\right) \left\{3 \left[4\left(\Gamma^2+\Phi^2\right) -\left(\Gamma^2-\Phi^2\right)^2\right] -8 \Gamma^2 \Phi^2\right\}  }{2 \Gamma^5 \Phi^5}\, \left[\alpha_1 - \beta_1\, \log\left(\frac{\Gamma+\Phi}{\Gamma-\phi} \right) \right] + \nonumber\\
    &&+\beta_1\,\frac{a^2 r \left[8 \left(\Gamma^2-\Phi^2\right)^2-3 \left(\Gamma^4-\Phi^4\right) \left(\Gamma^2-\Phi^2\right)+4 \left(\Gamma^2+\Phi^2\right)^2\right]}{\left(\Gamma^2-\Phi^2\right)\, \Gamma^4 \Phi^4 }\, ,\\
     D^{\xi}_{F,\parallel}&=&a\,\left\{\frac{ \left[8 \left(\Gamma^2+\Phi^2-2\right)-\left(\Gamma^2-\Phi^2\right)^2\right] \left(3 \Gamma^4+3 \Phi^4+2\Gamma^2 \Phi^2\right)}{8 \Gamma^5 \Phi^5}- \frac{4   \left(\Gamma^2+\Phi^2\right)}{\Gamma^3 \Phi^3}\right\}\times \nonumber \\
     &&\times  \left[\alpha_1 - \beta_1\, \log\left(\frac{\Gamma+\Phi}{\Gamma-\phi} \right) \right] +\nonumber \\ 
     &&+ \beta_1\, \frac{a \left[8 \left(\Gamma^2-\Phi^2\right)^2-3 \left(\Gamma^4-\Phi^4\right) \left(\Gamma^2-\Phi^2\right)+16 \left(\Gamma^2+\Phi^2\right)^2-48 \left(\Gamma^2+\Phi^2\right)\right]}{4 \Gamma^4 \Phi^4}\, ,
\end{eqnarray}

\begin{eqnarray}
     D^{r}_{F,\perp}&=& \frac{ 2 a \left(\Gamma^2-\Phi^2\right)  \left(\Gamma^2 \Phi^2-3 a^2 r^2 \left(\Gamma^2+\Phi^2\right)\right)}{\Gamma^5 \Phi^5}\, \left[\alpha_1 - \beta_1\, \log\left(\frac{\Gamma+\Phi}{\Gamma-\phi} \right) \right] + \nonumber \\
     &&+ 4\beta_1\, a\left[\frac{  \left(\Gamma^2+\Phi^2-4 a^2 r^2\right)}{\left(\Gamma^2-\Phi^2\right)\, \Gamma^2 \Phi^2 }-\frac{3 a^2 r^2  \left(\Gamma^2-\Phi^2\right)}{\Gamma^4 \Phi^4} \right]\, , \\
    D^{\phi}_{F,\perp}&=& -\frac{2 a^2 \, \left(\Gamma^2-\Phi^2\right)}{a r \, \Gamma^3\Phi^3}\, \left[\alpha_1 - \beta_1\, \log\left(\frac{\Gamma+\Phi}{\Gamma-\phi} \right) \right]  -\beta_1\, \frac{4 a^2 \left(\Gamma^2+\Phi^2\right)}{a r\,   \left(\Gamma^2-\Phi^2\right)\, \Gamma^2\Phi^2}\, , \\
      D^{\xi}_{F,\perp}&=&\frac{ a^2 r\,  \left[4 \left(2 \Gamma^2 \Phi^2+3 \Gamma^4+3 \Phi^4\right)-3 \left(\Gamma^2-\Phi^2\right) \left(\Gamma^4-\Phi^4\right)\right]}{2 \Gamma^5 \Phi^5}\, \left[\alpha_1 - \beta_1\, \log\left(\frac{\Gamma+\Phi}{\Gamma-\phi} \right) \right] + \nonumber \\
      &&+\beta_1\, \frac{ a^2 r \left\{3 \left[4\left(\Gamma^2+\Phi^2\right) -\left(\Gamma^2-\Phi^2\right)^2\right]-4\Gamma^2\Phi^2\right\}}{\Gamma^4\Phi^4}\, .
\end{eqnarray}

\subsection{Laboratory frame}

To present  the 1-form $A$ \eqref{udt} of the electromagnetic potential in the laboratory frame, it has to be pulled back to the laboratory frame via inverse transformation of \eqref{trans}, which implies:
\begin{eqnarray*}
    a\,  \dd t = \frac{\zeta\, \dd \tau - \tau \, \dd \zeta}{\zeta^2-\tau^2}\, .
\end{eqnarray*}
Fortunately, this transformation \eqref{trans} preserves $r$ and $\phi$ coordinates, therefore:
\begin{eqnarray}
    A  = \frac{U_{LAB} \, \left(\zeta\, \dd \tau - \tau \, \dd \zeta\right)}{a \left(\zeta^2-\tau^2\right)} \, ,
    \label{ALAB}
\end{eqnarray}
where $U_{LAB}$ denotes the potential function $U_F$ written in laboratory frame with cylindrical coordinates:
\begin{eqnarray*}
    U_{LAB}(r,\phi,\zeta, \tau):=U_F\left(r,\phi,\xi=\sqrt{\zeta^2-\tau^2}-\frac1a\right)\, .
\end{eqnarray*}
It implies, that all components of potential $U_F$ \eqref{UF}: the  radial function \eqref{RF} and polar parts of spherical harmonics \eqref{costh}, \eqref{sinth}, have to be transformed in the same manner. Derivation of electromagnetic fields $D_{LAB}$ and $H_{LAB}$ in laboratory frame reduces to calculating components of Faraday 2-form $f=\dd A$ \eqref{cfkl}, \eqref{cf0l}. The main difference between the co-moving  and the laboratory frames is the appearance of magnetic field $H_{LAB}$, what corresponds to the non-trivial part of 1-form $A$ \eqref{ALAB} proportional to $\dd\zeta$. Due to the fact that, obtained this way, formulae are long and complicated, the components $D_{LAB}^i$, $H_{LAB}^i$ are presented implicitly:
\begin{eqnarray}
    D^r_{LAB} &=&\frac{\zeta \, \partial_r U_{LAB}}{a\left(\zeta^2-\tau^2\right)}\, , \\
    D^{\phi}_{LAB} &=&\frac{a^3\zeta\, \partial_{\phi} U_{LAB}}{a^4 r^2(\zeta^2-\tau^2)}\, , \\
    D^{\zeta}_{LAB} &=&\frac{\zeta\, \partial_{\tau} U_{LAB} + \tau\, \partial_{\zeta} U_{LAB}  }{a\left(\zeta^2-\tau^2\right)}\, , \\
    H^r_{LAB} &=&-\frac{a^2\tau\, \partial_{\phi} U_{LAB}}{a^3 r\left(\zeta^2-\tau^2\right)}\, , \\
    H^{\phi}_{LAB} &=&\frac{a^2 \tau \, \partial_r U_{LAB}}{a^3r\, \left(\zeta^2-\tau^2\right)}\, , \\
    H^{\zeta}_{LAB} &=&0\, .
\end{eqnarray}

%\section*{Acknowledgements} 

\section*{APPENDIX}

\appendix

\section{Gegenbauer equation}
\label{gegapp}

The solutions of general Gegenbauer equation \eqref{gege} satisfy recurrence relations with respect to parameters $\alpha$ and $\lambda$ \cite{derez}, although, here are presented only those for $\lambda=1$ and $\alpha=l+\frac 12$ to generate the higher/lower multipoles:
\begin{eqnarray}
   G_{l+1}(z)&=& C_l\, \partial_z G_{l}(z)\, , \label{req1}\\
    G_{l-1}(z) &=& \tilde{C}_l\, \left[(1-z^2)\partial_z-\left(2l+1\right) z\right] G_{l}(z) \label{req2}\, ,
\end{eqnarray}
where $ C_l$, $\tilde{C}_l$ are constants, which are, in physical sense, irrelevant due to the fact, that the integral constants in the solution are determined by boundary Dirichlet conditions. Therefore, generating a solution for $l\geq 1$ is possible if it is known just one of them. Unfortunately, for $l=0$ the one branch of the solution is just a constant \eqref{l0} and second one is precisely $Z_0(z)$, what implies that generating  solutions for $l>0$ by recurrence formulae is impossible for the first branch (proportional to co the constant $\alpha_l$), whereas for the second one (proportional to the constant $\beta_l$) this procedure works well. Such an obstruction could be explicitly observed in the solution represented by hypergeometric series. However, this problem could be eliminated using the solution $Z_{l}(z)$ and the corresponding Wro\'nskian $W(z)$:
\begin{eqnarray}
    W(z) = \frac{(l+1)!}{(z^2-1)^{l+3/2}}\, .
    \label{wronskian}
\end{eqnarray}
Obtained this way family of functions $Q_l(z)$ again satisfy above recurrence relations \eqref{req1}, \eqref{req2} for $l\geq 1$.

\section*{Acknowledgements}
I want to thank prof. Jerzy Kijowski, prof. Jan Derezi{\'n}ski prof. Jacek Jezierski for many hours of fruitful discussions and help in editorial aspects of this paper.

\end{document}